\documentclass[preprint,showpacs,preprintnumbers,amsmath,amssymb]{revtex4}

\usepackage{graphicx}
\usepackage{dcolumn}
\usepackage{bm}

\begin{document}


\title{Fermionic zero modes in gauge and gravity backgrounds on $T^2$}

\author{Yishi Duan}
\affiliation{Institute of Theoretical Physics, Lanzhou University, Lanzhou 730000, China}
\author{Yuxiao Liu}
\thanks{Corresponding author}\email{liuyx01@st.lzu.edu.cn}\affiliation{Institute of Theoretical
Physics, Lanzhou University, Lanzhou 730000, China}
\author{Yongqiang Wang}
\affiliation{Institute of Modern Physics, Chinese Academy of Sciences, Lanzhou 730000,
China} \affiliation{Institute of Theoretical Physics, Lanzhou University, Lanzhou 730000,
China}
\date{\today}

\begin{abstract}
In this note we study fermionic zero modes in gauge and gravity
backgrounds taking a two dimensional compact manifold $T^2$ as
extra dimensions. The result is that there exist massless Dirac
fermions which have normalizable zero modes under quite general
assumptions about these backgrounds on the bulk. Several special
cases of gauge background on the torus are discussed and some
simple fermionic zero modes are obtained.
\end{abstract}

\maketitle

\section{Introduction}
Recently, there has been active interest in the brane world scenarios[1-8]. The
pioneering work was done by Randall and Sundrum \cite{RS1,RS2}. In a five-dimensional
brane world scenario, they assumed that the bulk geometry is warped. The RSI scenario
provides a way to solve the hierarchy problem, and the RSII scenario gives Newton's law
of gravity on the brane of positive tension embedded in an infinite extra dimension.
Another interesting topic is localization of different fields on a brane. It has been
shown that the graviton \cite{RS2} and the massless scalar field \cite{B.Bajc} have
normalizable zero modes on branes of different types, that the Abelian vector fields are
not localized in the RS model in five dimensions but can be localized in some
higher-dimensional generalizations of it \cite{I.Oda}. In Ref. \cite{S.Randjbar-Daemi, LiuyuxiaoMPLA2005} it
was shown that there exist massless Dirac fermions under quite general assumptions about
the geometry and topology of the internal manifold of the higher-dimensional warp factor
compactification. However, these fermionic modes are generically non-normalizable. To
avoid this problem, the authors include a Yukawa-type coupling to a scalar field of a
domain-wall type. While we shall study fermionic zero modes in gauge and gravity
backgrounds taking a two dimensional compact manifold $T^2$ as extra dimensions.

The paper is organized as follows: In section II, we first present Dirac equation and the
effective Lagrangian in gauge and gravity backgrounds. In section III, we discuss and
solve the fermionic zero modes under several simple assumptions for gauge field on the
torus. In the last section, a brief conclusion is presented.

\section{Dirac equation in gauge and gravity backgrounds}
We shall consider a six-dimensional space-time manifold
$M^{4}\times T^{2}$ with the geometry
\begin{equation} \label{Metric}
ds^2=G_{MN}dx^M dx^N = \eta_{\mu \nu} dx^\mu dx^\nu - (R_1 + R_2
\cos\varphi)^2 d\theta^2 - R_2^2 d\varphi^2,
\end{equation}
where $\eta_{\mu \nu}$ is the metric of our four-dimensional
Minkowski space-time $M^4$, capital Latin indices
$M,N=0,\cdots,5$, Greek indices $\mu,\nu=0,\cdots,3$, $x^4=\theta$
and $x^5=\varphi$ are coordinates on $T^2=S^1 \times S^1$, $R_1$
and $R_2$ are radii of the two circles $S^1$, respectively. The
six-dimensional Dirac equation is
\begin{equation} \label{DiracEq1}
\Gamma^A E^{M}_{A} (\partial_M - \Omega_M  + i A_M)
\Psi(x,\theta,\varphi)=0,
\end{equation}
where $E^{M}_{A}$ is the {\sl sechsbein} with
\begin{equation} \label{sechsbein}
E^{M}_{A} = (\delta^{\mu}_{A},\frac{1}{R_1+R_2
\cos\varphi}\delta^{4}_{A},\frac{1}{R_2}\delta^{5}_{A}),
\end{equation}
and capital Latin indices $A,B=0,\cdots,5$ correspond to the flat
tangent six-dimensional Minkowski space-time,
$\Omega_M=\frac{1}{2} \Omega_M^{AB}I_{AB}$ is the spin connection
with the following representation of six-dimensional 8 $\times$ 8
Dirac matrices $\Gamma^A$:
\begin{equation}
\Gamma^A=\begin{pmatrix}
  0 & \Sigma^A \\
  \bar{\Sigma}^A & 0 \\
\end{pmatrix}, \nonumber
\end{equation}
where $\Sigma^0 = \bar{\Sigma}^0 = \gamma^0
\gamma^0$; $\Sigma^i = -\bar{\Sigma}^i = \gamma^0 \gamma^i$;
$\Sigma^4 = -\bar{\Sigma}^4 = i\gamma^0 \gamma^5$; $\Sigma^5 =
-\bar{\Sigma}^5 = \gamma^0$, $\gamma^{\mu}$ and $\gamma^5$ are
usual four-dimensional Dirac matrices in the chiral
representation:
\begin{equation}
\gamma^0=\begin{pmatrix}
  0 & 1 \\
  1 & 0 \\
\end{pmatrix},\;\;\;
\gamma^i=\begin{pmatrix}
  0 & \sigma^i \\
  -\sigma^i & 0 \\
\end{pmatrix},\;\;\;
\gamma^5=i \gamma^0\gamma^1\gamma^2\gamma^3=\begin{pmatrix}
  1 & 0 \\
  0 & -1 \\
\end{pmatrix}, \nonumber
\end{equation}
here $\sigma^i$ are the Pauli matrices. $\Gamma^A$ have the
following relation:
\begin{equation}
\Gamma^A\Gamma^B+\Gamma^B\Gamma^A=2\eta^{AB}I, \nonumber
\end{equation}
in which $\eta^{AB}=$ diag$ (+,-,\cdots,-)$ is the six-dimensional
Minkowski metric.

The non-vanishing component of $\Omega_M$ is
\begin{equation}
\Omega_{4}=\frac{1}{2} \sin\varphi \Gamma^4 \Gamma^5.
\end{equation}
Assume $A_{\mu}=A_{\mu}(x)$, $A_{\theta}=A_{\theta}(\theta,\varphi)$,
$A_{\varphi}=A_{\varphi}(\theta,\varphi)$. The six-dimensional Dirac equation
(\ref{DiracEq1}) then becomes
\begin{equation} \label{DiracEq2}
\left \{\Gamma^a \delta^{\mu}_{a} (\partial_{\mu} + i A_{\mu}) + \frac{\Gamma^4
(\partial_{\theta} + i A_{\theta})}{R_1+R_2 \cos\varphi}  + \frac{\Gamma^5}{R_2} \left
(\partial_{\varphi} + \frac{R_2 \sin\varphi}{2(R_1+R_2 \cos\varphi)} + i A_{\varphi}
\right ) \right \} \Psi = 0.
\end{equation}
We denote the Dirac operators on $M^4$ and $T^2$ with $D_M$ and
$D_T$, respectively:
\begin{eqnarray}
D_M &=& \bar{\Gamma} \Gamma^a \delta^{\mu}_{a} (\partial_{\mu} + i A_{\mu}),\\
D_T &=& \frac{\bar{\Gamma}\Gamma^4 (\partial_{\theta} + i
A_{\theta})}{R_1+R_2 \cos\varphi}
 + \frac{\bar{\Gamma}\Gamma^5}{R_2} \left ( \partial_{\varphi} +
\frac{R_2 \sin\varphi}{2(R_1+R_2 \cos\varphi)}+i A_{\varphi}
\right ),
\end{eqnarray}
where $\bar{\Gamma}=\Gamma^0 \Gamma ^1 \Gamma ^2 \Gamma ^3$. Then
we have the following commutative relations:
\begin{equation}
[D_M,D_T]=0, \nonumber
\end{equation}
and can expand any spinor $\Psi$ in a set of eigenvectors $\phi_m$
of the operator $D_T$
\begin{equation}
D_T\phi_m=\lambda_m\phi_m.
\end{equation}
There may exist a set of discrete eigenvalues $\lambda_m$ with
some separation. All these eigenvalues play a role of the mass of
the corresponding four-dimensional excitations \cite{M.V.Libanov}.
We assume that the energy scales probed by a four-dimensional
observer are smaller than the separation, and thus even the first
non-zero level is not excited. So, we are interested only in the
zero modes of $D_T$:
\begin{equation} \label{DiracEqOnT2}
D_T\phi=0.
\end{equation}
This is just the Dirac equation on the torus $T^2$ with gauge and
gravity backgrounds. For fermionic zero modes, we can write
\begin{equation}
\Psi(x,\theta,\varphi)=\psi(x) \phi(\theta,\varphi),
\end{equation}
where $\psi$ and $\phi$ satisfy
\begin{eqnarray}
D_M \psi(x) &=& 0,  \\
D_T \phi(\theta,\varphi) &=& 0.
\end{eqnarray}
The effective Lagrangian for $\psi$ then becomes
\begin{eqnarray}
&\int& d\theta d\varphi \sqrt{-G}
\bar{\Psi} \Gamma^A E^{M}_{A} (\partial_M - \Omega_M + i A_M) \Psi  \nonumber \\
&=& \bar{\psi} \Gamma^a \delta^{\mu}_{a} (\partial_{\mu}  + i A_{\mu}) \psi \times  \int
d\theta d\varphi (R_1+R_2 \cos\varphi)R_2 \phi^\dag \phi.
\end{eqnarray}
Thus, to have the localization of gravity and finite kinetic energy for $\psi$, the above
integral must be finite. This can be achieved if the function $\phi(\theta,\varphi)$ does
not diverge on the whole torus. This is very different from the case considered by
Randjbar-Daemi and Shaposhnikov \cite{S.Randjbar-Daemi}, who did this work on a $D=D_1 +
D_2 +1$ dimensional manifold with the geometry
\begin{equation}
ds^2=e^{A(r)}\eta_{\mu\nu}dx^{\mu}dx^{\nu} +
e^{B(r)}g_{mn}dy^m dy^n + dr^2,
\end{equation}
under the assumptions $A_{\mu}=A_r=0$ and $A_m=A_m(y)$. Their
effective Lagrangian for $\psi$ is
\begin{eqnarray}
&\int& dr dy \sqrt{-G} \bar{\Psi} \Gamma^A E^{M}_{A} (\partial_M - \Omega_M + A_M) \Psi \nonumber\\
&=& \bar{\psi} \Gamma^a \delta^{\mu}_{a} \partial_{\mu} \psi \times \int dr \;
e^{-A(r)/2} \int dy \sqrt{\det(g_{mn})} \phi^\dag \phi,
\end{eqnarray}
and the difficulty is that the integral $\int dr e^{-A(r)/2}$ is
infinite for the exponential warp-factor $A \propto -|r|$
considered in the literature so far.\\

\section{Fermionic zero modes on $T^2$}
In this section, we discuss and solve the fermionic zero modes
under some simple assumptions for gauge field
$A_{\theta}$ and $A_{\varphi}$ on the torus.\\

\subsection{Case I: $A_{\theta}=A_{\theta}(\theta)$, $A_{\varphi}=A_{\varphi}(\varphi)$}
In this assumption, the Dirac equation (\ref{DiracEqOnT2}) becomes
\begin{equation} \label{DiracEqCaseI}
\left \{ \frac{\bar{\Gamma}\Gamma^4 (\partial_{\theta} + i
A_{\theta}(\theta))}{R_1+R_2 \cos\varphi}  +
\frac{\bar{\Gamma}\Gamma^5}{R_2} \left (\partial_{\varphi} +
\frac{R_2 \sin\varphi}{2(R_1+R_2 \cos\varphi)}+i
A_{\varphi}(\varphi) \right ) \right \} \phi = 0.
\end{equation}
Here $\phi$ can be written as the following form:
\begin{equation}
\phi (\theta,\varphi)=f(\theta) h(\varphi),
\end{equation}
where $f$ and $h$ satisfy
\begin{eqnarray}
\left ( \partial_{\theta} + i A_{\theta}(\theta) \right )
f(\theta) &=& 0,\\
\left (\partial_{\varphi} + \frac{R_2 \sin\varphi}{2(R_1+R_2
\cos\varphi)}+i A_{\varphi}(\varphi) \right ) h(\varphi) &=& 0.
\end{eqnarray}
Solve the last two equations one can easily get the formalized
solutions:
\begin{eqnarray}
f(\theta) &=& e^{- i \int{d \theta  A_{\theta}(\theta)}},\\
h(\varphi) &=& \sqrt{R_1+R_2 \cos\varphi} \;\; e^{-i \int{d
\varphi
A_{\varphi}(\varphi) }},\\
\phi(\theta,\varphi) &=& \sqrt{R_1+R_2 \cos\varphi} \;\; e^{-i
\int{d \theta A_{\theta}(\theta)}-i \int{d \varphi
A_{\varphi}(\varphi) }}.
\end{eqnarray}
The effective Lagrangian for $\psi$ then becomes
\begin{eqnarray}
&\int& d\theta d\varphi \sqrt{-G}
\bar{\Psi} \Gamma^A E^{M}_{A} (\partial_M - \Omega_M + i A_M) \Psi \nonumber\\
&=& \bar{\psi} \Gamma^a \delta^{\mu}_{a} (\partial_{\mu}  + i A_{\mu}) \psi
\int^{2\pi}_{0} d\theta
 \| e^{-i \int{d \theta A_{\theta}(\theta)}}
\|^2 \nonumber\\
&\times& \int^{2\pi}_{0}{d\varphi} (R_1+R_2 \cos\varphi)^{2}
R_{2}\| e^{-i \int{d \varphi A_{\varphi}(\varphi)}} \|^2.\nonumber\\
&=& \bar{\psi} \Gamma^a \delta^{\mu}_{a} (\partial_{\mu} + i A_{\mu}) \psi  \times 2
\pi^2 (2 R_1^2 + R_2^2)R_2 .
\end{eqnarray}
This result shows that, whatever the forms of $A_{\theta}(\theta)$ and
$A_{\varphi}(\varphi)$ are, the effective Lagrangian for $\psi$ has the same form.

\subsection{ Case II: $A_{\theta}=A(\varphi)$, $A_{\varphi}=0$}
This simple case is considered by J. M. Frere etc.
\cite{J.-M.Frere}. But a complex scalar field
$\Phi(\theta,\varphi)$ on $S^2$ is included in their work. For our
current ansatz, The Dirac equation (\ref{DiracEqOnT2}) is read as
\begin{equation} \label{DiracEqCaseII}
\left \{ \frac{\bar{\Gamma}\Gamma^4 \partial_{\theta}}{R_1+R_2
\cos\varphi} + \frac{\bar{\Gamma}\Gamma^5}{R_2} \left
(\partial_{\varphi} + \frac{R_2 \sin\varphi + 2iR_{2}
\Gamma^4\Gamma^5 A(\varphi)}{2(R_1+R_2 \cos\varphi)}\right )
\right \} \phi = 0
\end{equation}
with
\begin{equation}
\phi(\theta,\varphi)=f(\theta)h(\varphi),
\end{equation}
where $f(\theta)=C$ is a constant and $h$ satisfies
\begin{equation}
\left \{\partial_{\varphi} + \frac{R_2 \sin\varphi + 2iR_{2}
\Gamma^4\Gamma^5 A(\varphi)}{2(R_1+R_2 \cos\varphi)}\right \}
h(\varphi)=0,
\end{equation}
i.e.
\begin{equation}
\left \{\partial_{\varphi} + \frac{R_2 \sin\varphi - 2 R_{2}
\gamma^5 A(\varphi)}{2(R_1+R_2 \cos\varphi)}\right \}
h(\varphi)=0.
\end{equation}
Imposing the chirality condition $\gamma^5 h(\varphi) = +h
(\varphi)$, the above equation can be written as
\begin{equation}
\left \{\partial_{\varphi} + \frac{R_2 \sin\varphi}{2(R_1+R_2
\cos\varphi)} - \frac{ R_{2} A(\varphi)}{R_1+R_2
\cos\varphi}\right \} h(\varphi)=0.
\end{equation}
So, the formalized solution of $h$ can be given out
\begin{equation}
h(\varphi) =\sqrt{R_1+R_2 \cos\varphi} \;\; e^{ \int{d\varphi
\frac{R_2 A(\varphi)}{R_1+R_2 \cos\varphi}}}
\left(%
\begin{array}{c}
  1 \\
  0 \\
\end{array}%
\right).
\end{equation}
Let $C=1$, the effective Lagrangian for $\psi$ then becomes
\begin{eqnarray}
&\int& d\theta d\varphi \sqrt{-G}
\bar{\Psi} \Gamma^A E^{M}_{A} (\partial_M - \Omega_M + i A_M) \Psi \nonumber\\
&=& \bar{\psi} \Gamma^a \delta^{\mu}_{a} (\partial_{\mu} + i A_{\mu}) \psi
\int^{2\pi}_{0} {d\theta}\int^{2\pi}_{0} {d\varphi (R_1+R_2 \cos\varphi)^2 R_{2} \| e^{
\int{d\varphi \frac{ R_2 A(\varphi)}{R_1+R_2 \cos\varphi}}} \|^2}.
\end{eqnarray}
When the Yang-Mills field on $T^2$ disappeared, the zero mode
$\phi(\theta, \varphi)$ take the following simplest form
\begin{equation}
\phi(\theta, \varphi)=\sqrt{R_1+R_2 \cos\varphi}
\left(%
\begin{array}{c}
  1 \\
  0 \\
\end{array}%
\right),
\end{equation}\\
and the effective Lagrangian is
\begin{eqnarray}
&\int& d\theta d\varphi \sqrt{-G}
\bar{\Psi} \Gamma^A E^{M}_{A} (\partial_M - \Omega_M + i A_M) \Psi \nonumber\\
&=& \bar{\psi} \Gamma^a \delta^{\mu}_{a} (\partial_{\mu} + i A_{\mu}) \psi  \times 2
\pi^2 (2 R_1^2 + R_2^2)R_2 .
\end{eqnarray}

Now we come to the issue of the presence of the zero modes. The the number of zero-modes of the Dirac operator is decided by the index of it. The index of the Dirac operator on manifold $K$ is defined as the difference $n_{+} - n_{-}$ between the number $n_{+}$ of right-handed four-dimensional fermions obtained by dimensional reduction and the number $n_{-}$ of left-handed 4D fermions. This number is a topological quantity of the manifold upon compactification and the gauge bundles the Dirac operator might be coupled to. Indeed, this index can be computed in terms of characteristic classes of the tangent and gauge bundles. The Atiyah--Singer index theorem in two dimensions gives the difference \cite{Atiyah1968_485, Atiyah1968_546, Atiyah1973_279}
\begin{equation}
n_{+} - n_{-}=\frac{e}{4\pi} \int_\mathcal{M} d^2 q \,
\varepsilon^{\mu\nu} F_{\mu\nu} \,,
\end{equation}
where $\varepsilon^{\mu\nu}$ $(\varepsilon^{12}=1)$ is the contravariant Levi--Civita tensor {\em density} in two dimensions, and $F_{\mu\nu}$ is the field strength of $A_{\mu}$,
\begin{equation}
F_{\mu\nu}\equiv \partial_{\mu}A_{\nu}-\partial_{\nu}A_{\mu}
-ie[A_{\mu}, A_{\nu}] \,.
\end{equation}
If we take $K=S^2$ with a U(1) magnetic monopole field of charge $n$ on it, the number of chiral families will then be equal to $n$ \cite{Randjbar1983, Deguchi2005}. Here, we can consider zero modes of the Dirac operator in the background of Abelian gauge potentials representing Dirac strings and center vortices on the torus $T^2$. The result is for a two-vortex gauge potential (smeared out vortices) there is one normalizable zero mode which has exactly one zero on the torus \cite{Reinhardt2002}. The probability density of the spinor field is peaked at the positions of the vortices.

\section{Conclusion}
In conclusion, we studied fermionic zero modes and effective Lagrangian in gauge and
gravity backgrounds taking a two dimensional compact manifold $T^2$ as extra dimensions.
The result is that there exist massless Dirac fermions under quite general assumptions
about the gauge field on the bulk. These fermionic zero modes are generically
normalizable and we need not include any other field. Furthermore, Several special cases
of these backgrounds are also discussed. Especially, in the case of $A_{\theta} =
A_{\theta}(\theta)$, $A_{\varphi} = A_{\varphi}(\varphi)$, we got very simple fermionic
zero mode on the torus and the effective Lagrangian for $\psi$, which has the same form
whatever these gauge fields are.

\section{Acknowledgements}
It is a pleasure to thank Dr. Lijie Zhang and Zhenhua Zhao
for discussions. This work was supported by the National
Natural Science Foundation and the Doctor Education Fund of
Educational Department of the People's Republic of China.

\end{document}